\documentclass{osa-article}

\journal{oe}

\usepackage{verbatim}
\usepackage{multirow}
\usepackage{booktabs}
\usepackage{graphicx}

\begin{document}

\title{Atomic Ramsey interferometry with S- and D-band in a triangular optical lattice}

\author{Xiangyu Dong,\authormark{1} Chengyang Wu,\authormark{1} Zhongcheng Yu,\authormark{1} Jinyuan Tian,\authormark{1} Zhongkai Wang,\authormark{2} Xuzong Chen,\authormark{1} Shengjie Jin,\authormark{1,3} and Xiaoji Zhou\authormark{1,4}}

\address{\authormark{1}State Key Laboratory of Advanced Optical Communication System and Network, School of Electronics, Peking University, Beijing 100871, China\\
\authormark{2}Shenzhen Institute for Quantum Science and Engineering, Southern University of Science and Technology, Shenzhen 518055, China.\\
\authormark{3}jinshengjie@pku.edu.cn\\
\authormark{4}xjzhou@pku.edu.cn}

\begin{abstract}

Ramsey interferometers have wide applications in science and engineering. Compared with the traditional interferometer based on internal states, the interferometer with external quantum states has advantages in some applications for quantum simulation and precision measurement. Here, we develop a Ramsey interferometry with Bloch states in S- and D-band of a triangular optical lattice for the first time. The key to realizing this interferometer in two-dimensionally coupled lattice is that we use the shortcut method to construct $\pi/2$ pulse. We observe clear Ramsey fringes and analyze the decoherence mechanism of fringes. Further, we design an echo $\pi$ pulse between S- and D-band, which significantly improves the coherence time. This Ramsey interferometer in the dimensionally coupled lattice has potential applications in the quantum simulations of topological physics, frustrated effects, and motional qubits manipulation.
\end{abstract}

\section{Introduction}

Proposed in the late 1940s\cite{PhysRev.76.996}, the Ramsey interferometer (RI) has adapted itself to the recent advances in the field of precision measurement and quantum information for its higher resolution, such as in spectroscopy\cite{Kandula:08,Wang:22,PhysRevLett.90.143602}, quantum gate\cite{Reddy:18,PhysRevResearch.4.023071,science.1231675,Liu:13}, atomic clock\cite{PhysRevLett.82.4619, PhysRevLett.106.130801}, gyroscope\cite{PhysRevLett.124.120403, PhysRevLett.97.240801}, remote entanglement\cite{Dumur2021}, and quantum simulations\cite{PhysRevX.2.041020, science.aaf5134, atoms9020022}. Typical sequences of RI consist of two $\pi/2$ pulses\cite{Hu2018}, and the key to realizing the interferometer is to select appropriate quantum states for interference and construct corresponding $\pi/2$ pulses. Conventional RIs are based on the internal states of atoms, which have been widely used in accurate quantum state engineering and quantum metrology\cite{science.1231675,PhysRevLett.82.4619,PhysRevLett.79.769,science.aaf5134}. The interferometers using external states, for example, the motional states of a Bose-Einstein condensate (BEC), have emerged in recent years\cite{vanFrank2014, Hu2018}. These RIs display advantages in the study of multi-body physics and the manipulation of motional qubits\cite{vanFrank2014, Hu2018,PhysRevA.104.L060601}, such as the investigation of physics decoherence and relaxation dynamics in a doping system\cite{PhysRevLett.111.070401} and the measurement of the phononic Lamb shift\cite{PhysRevX.6.041041}.

Ultracold atoms in optical lattices are a powerful platform for precision measurement and quantum information. Over the last few years, considerable attention has been paid to preparing atoms into different bands of optical lattices, which is used for the quantum simulation of orbital physics\cite{Wirth2011,Li_2016,Zhou_2018,PhysRevLett.121.265301,PhysRevLett.126.035301,Wang2021,PhysRevLett.111.205302}, and quantum sensing\cite{PhysRevLett.124.120403,2208.05368}. Combining the RI with the BEC populated in different bands of optical lattices is expected to be applied to quantum computation, quantum simulation and quantum sensing. In our previous work, a RI  with trapped motional quantum states in the S- and D-band of a one-dimensional (1D) optical lattice is constructed by a shortcut method, and has been used to realize an atom-orbital qubit\cite{Hu2018,PhysRevA.104.L060601}. More interesting physics, such as frustrating effects\cite{science.1114727,Eckardt_2010} and topological physics\cite{Jotzu2014}, requires two-dimensional (2D) dimensionally-coupled lattices, for example, a triangular optical lattice\cite{Jin_2019,PhysRevA.104.033326}. These 2D systems can also be used to find novel quantum phases\cite{PhysRevLett.95.127205, PhysRevLett.95.127207, PhysRevLett.97.190406,PhysRevLett.126.035301,Wang2021} and realize long-lived qubits\cite{Hartke2022}. However, despite these advantages, the development of RI in 2D dimensionally-coupled lattices remains hindered by difficulties in the experimental manipulation of atoms in these lattices. The key point is that the $\pi/2$ pulse for RI needs to be accurately designed in the case of a more complex band structure due to dimensional coupling.

In this paper, we develop a RI with the Bloch states in the S- and D-band of a 2D triangular optical lattice for the first time. Using a shortcut method, we construct a shortcut $\pi/2$ pulse for manipulating the BEC in this RI, which requires more complex numerical calculations than the 1D case. Then we manage to observe clear time-dependent Ramsey fringes and analyze the decay mechanism of the fringes in experiments and theory, which is mainly caused by the quasi-momentum distribution of atoms. Furthermore, we design a band echo $\pi$ pulse for the Bloch states in the S- and D-band, which tremendously improves the coherence time of the RI.

\begin{figure}[!htbp]
	\centering
	\includegraphics[width=0.8\linewidth]{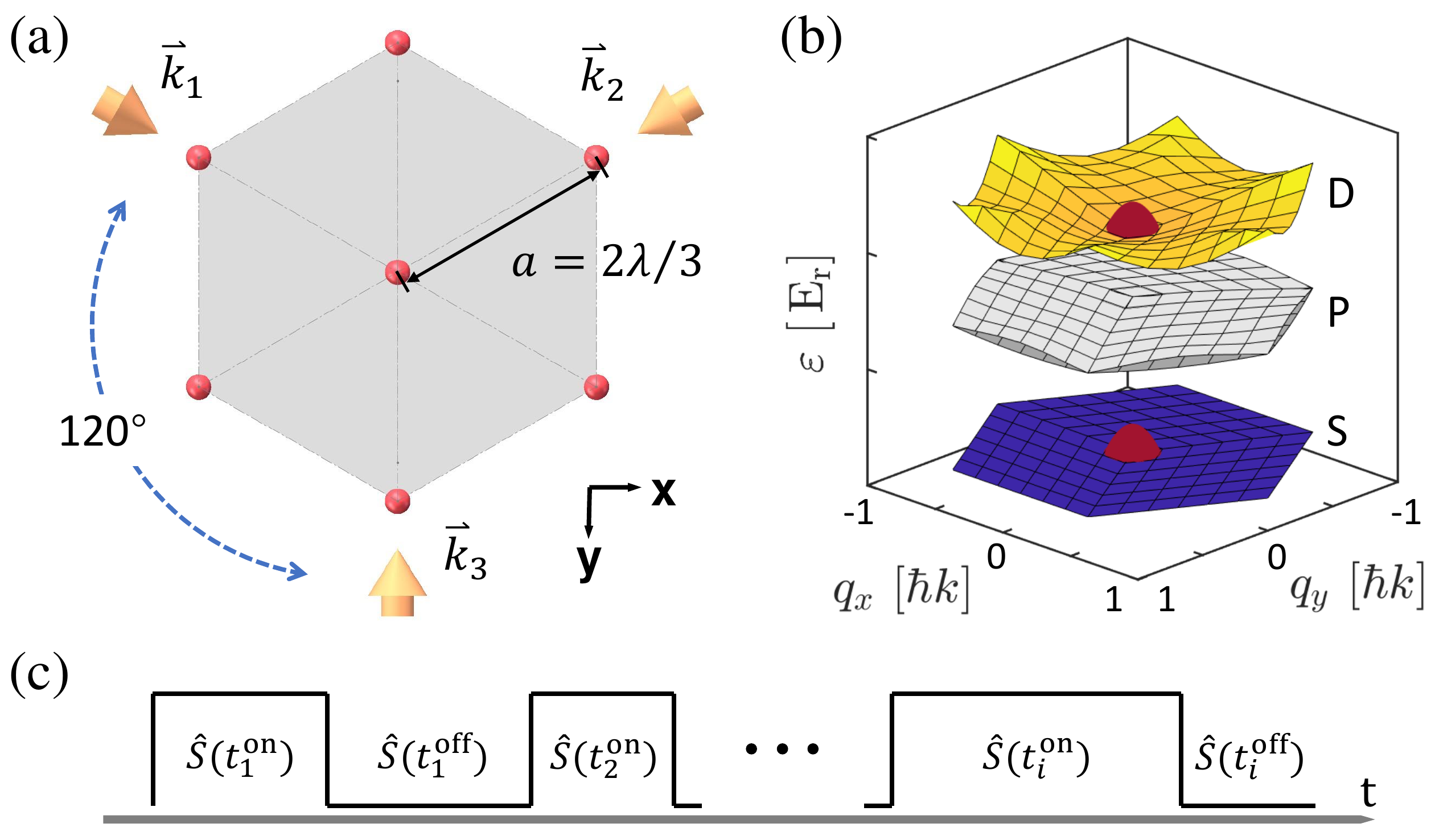}
	\caption{(a) Formation of a 2D triangular optical lattice. The yellow arrows in the \textit{x-y} plane stand for the lattice beams with wave vectors $\vec{k}_j,\,j=1, 2, 3$. They converge at one point with an enclosing angle of $120^{\circ}$. (b) Band structure of the triangular lattice. The blue and yellow curved surface represent the S- and D-band, respectively. The two grey surfaces in the middle are the P-bands. (c) Shortcut sequence. Each on/off pulse has its own evolution operator $\hat{S}(t_i^\mathrm{on})$ or $\hat{S}(t_i^\mathrm{off})$, where $t$ is the duration time and $i=1,\,2,\,...$}
	\label{fig:fig1}
\end{figure}\textbf{}

\section{Construction for the Ramsey interferometer in a triangular optical lattice}

We choose to construct RI in triangular lattice because of its advantages in topology physics, frustrated effects and other novel physical research. As shown in Fig.\ref{fig:fig1}(a), this lattice is made up of three travelling beams denoted as the wavevectors $\vec{k}_j,\,j=1, 2, 3$ and $\vec{k}_1+\vec{k}_2=-\vec{k}_3$ with $|\vec{k}_j|=2\pi/\lambda$, where $\lambda$ is the wavelength. The three beams intersect at an enclosing angle of $120^{\circ}$, which are linearly polarized perpendicular to the lattice plane (\textit{x-y} plane). The band structure of the lowest four bands of the triangular lattice is plotted in Fig.\ref{fig:fig1}(b). The first and fourth bands are S- and D-bands, and the middle bands are two P-bands. We choose the Bloch states at $\Gamma$ points (where quasi-momentum $q_{x,\,y}=0$) of the S-band (marked as $\left|\psi_S\right>$) and the D-band (marked as $\left|\psi_D\right>$) as the two states for interference to conduct our RI. There are two advantages to selecting these two states for interference in this way. The first is that the parity of the two states in S- and D-band are the same even parity, which is different from that of the P-band (odd parity). The same parity makes the operation process in RI simpler and shorter, because there is no need to change the parity\cite{PhysRevLett.121.265301,Zhou_2018}. Meanwhile, the same parity also avoids atomic leakage to the P-band during RI\cite{PhysRevA.104.L060601}. The second advantage is that the $\Gamma$ point is the lowest point of the S- or D-band, which makes the quantum state more stable, leading to a longer controllable time\cite{PhysRevA.104.033326}.

The key to constructing RI with Bloch bands in 2D optical lattice is to obtain the effective $\uppi/2$ pulse. There are three challenges in designing the  $\uppi/2$ pulse. The first is the dimensional coupling effect\cite{PhysRevA.104.033326} in the triangular lattice caused by the non-orthogonal laser paths. We should transmit the atoms to the target state in both dimensions \textit{x, y} simultaneously, which significantly increases the complexity of the pulse design. The second is the multi-target design. A $\uppi/2$ pulse should be able to rotate any quantum state on the Bloch sphere (or any superposition states of $\left|\psi_D\right>$ and $\left|\psi_S\right>$) by $\uppi/2$. The third challenge is the rapid manipulation of the quantum states. Despite we choose the states $\left|\psi_S\right>$ and $\left|\psi_D\right>$ with longer lifetime, the coherence time of these states will also limit the time for manipulation\cite{Hu2018,PhysRevA.104.L060601}.

To overcome these challenges, we use the shortcut method to construct the $\uppi/2$ pulse. Unlike the adiabatic control method\cite{Greiner2002,PhysRevLett.126.035301}, the shortcut is a rapid nonadiabatic control method for manipulating quantum states in optical lattices with high fidelity\cite{Zhou_2018,Hu2018,PhysRevA.104.L060601,PhysRevLett.121.265301,PhysRevA.104.013309,PhysRevA.96.043609,PhysRevLett.127.190605}. This shortcut process consists of several pulses shown in Fig.\ref{fig:fig1}(c), which can be completed within two hundred microseconds\cite{Zhou_2018}. In each pulse, the evolution operators $\hat{S}(t_i^{\rm{on}})$ (for lattice on) and $\hat{S}(t_i^{\rm{off}})$ (for lattice off) change the population of atoms in different bands and the phase of the quantum states depending on the time durations $t_i^{\rm{on}}$ and $t_i^{\rm{off}}$. By selecting appropriate sequence ($t_i^{\rm{on}},t_i^{\rm{off}}$), we can achieve the required manipulation of quantum states.

For the shortcut $\uppi/2$ pulse in our RI, the evolution operator can be expressed as $\hat{R}(\uppi/2)=\sum_i{\hat{S}(t_i^{\rm{off}})\hat{S}(t_i^{\rm{on}})}$. After the shortcut pulse, the two states $\left|\psi_S\right>$ and $\left|\psi_D\right>$ are transformed into $\psi_{f1}=\hat{R}(\uppi/2)\left|\psi_S\right>$ and $\psi_{f2}=\hat{R}(\uppi/2)\left|\psi_D\right>$, respectively. When satisfying two conditions $\psi_{f1}=\left|\psi_{m1}\right>\equiv(\left|\psi_S\right>+\left|\psi_D\right>)/\sqrt{2}$ and  $\psi_{f2}=\left|\psi_{m2}\right>\equiv(\left|\psi_D\right>-\left|\psi_S\right>)/\sqrt{2}$, the operation $\hat{R}(\uppi/2)$ can be considered as a $\uppi/2$ pulse, which can realize the $\uppi/2$ rotation on any superposition states of $\left|\psi_S\right>$ and $\left|\psi_D\right>$\cite{Hu2018}. Hence, we define the fidelity of the $\uppi/2$ pulse as $\eta=\left|\left<\psi_{f1}|\psi_{m1}\right>+\left<\psi_{f2}|\psi_{m2}\right>\right|/2$. With the numerical calculation, we optimize the pulse sequence $(t^{\rm{on}}_i,t^{\rm{off}}_i)$ by gradient descent method so that the fidelity approaches 100\%. Then we obtain the $\uppi/2$ pulse sequence in the triangular lattice (see the next section for specific time sequence).

Based on this elaborate $\uppi/2$ pulse, the process of the RI in a triangular lattice are as follows. Atoms are prepared to the initial state $\psi_i=\left|\psi_S\right>$ in the beginning, followed by the $\uppi/2$ pulse to excite atoms into the superposition states ($\left|\psi_S\right>$ and $\left|\psi_D\right>$). The lattice will remain on for a holding time $t_\mathrm{OL}$ after the first $\uppi/2$ pulse, thus atoms populated in different bands evolve in the lattice potential and accumulate phase separately ($\phi_S$ for $\left|\psi_S\right>$ and $\phi_D$ for $\left|\psi_D\right>$). The phase difference is considered to be $\phi_D-\phi_S=\Delta\epsilon t_\mathrm{OL}/\hbar$ where $\Delta\epsilon$ is the band gap between S- and D- band at $\Gamma$ points. After the second $\uppi/2$ pulse atoms with different phases accumulated interfere. The final state can be written as
\begin{equation}\label{eq:pi2}
\begin{split}
    \psi _{f} &= \hat{R}(\uppi/2)\hat{U}(t_\mathrm{OL})\hat{R}(\uppi/2)\psi_i \\
&= [(e^{i\phi_S}-e^{i\phi_D})\left|\psi_S\right>+(e^{i\phi_S}+e^{i\phi_D})\left|\psi_D\right>]/2,
\end{split}
\end{equation}
where $\hat{U}(t_\mathrm{OL})$ is the evolution operator for the lattice holding stage. From Eq.~(\ref{eq:pi2}), the relationship between population of atoms in D-band (denoted as $P_D$) and the holding time $t_\mathrm{OL}$ is
\begin{equation}\label{eq:ramsey}
		p_D(t_\mathrm{OL})=(1+\cos(\Delta\epsilon t_\mathrm{OL}/\hbar))/2,
\end{equation}
which indicates that the population of atoms in D-band oscillates with holding time $t_\mathrm{OL}$, that is, the Ramsey fringes in our RI.

\section{Experiment description}

The $^{87}\rm{Rb}$ BECs are prepared in a hybrid  optical-magnetic trap, which contains a 1064 nm laser beam along the \textit{x}-direction (the x-direction is shown in Fig.\ref{fig:fig1}(a)), providing an optical dipole trap, and a gradient magnetic field\cite{PhysRevLett.121.265301, PhysRevLett.126.035301, PhysRevA.101.013612}. With harmonic trapping frequencies of $(\omega_x,\,\omega_y,\,\omega_z)=2\uppi\times(28,\,55,\,60)\,\mathrm{Hz}$, we obtain a nearly pure BEC with $2\times10^5$ atoms populated on the $\left|F=2, m_F=+2\right>$ state. The wavelength of the triangular optical lattice is $\lambda=1064\,\mathrm{nm}$ and the lattice depth is $V_{\mathrm{OL}}=5\,\mathrm{E_r}$, where $\mathrm{E_r}=\hbar^2k^2/2m$, with $m$ the mass of the atom. The depth we choose is suitable for the RI. For a very deep depth, the coherence of atoms between the different sites becomes weak, which will affect the amplitude of the interference fringes. With a lattice trap that is too shallow, it is difficult to construct the $\uppi/2$ pulse, because the band gap (between S- and D-band at zero quasi-momentum) is very small and the manipulation time is longer. For $V_{\mathrm{OL}}=5\,\mathrm{E_r}$, we calculate the time sequence of the shortcut $\uppi/2$ pulse \{$(t_1^{\rm{on}},t_1^{\rm{off}},...,t_5^{\rm{on}},t_5^{\rm{off}})=(2.7,16.8,21.9,7.9,25.0,6.8,25.3,14.2,11.0,16.1)\,\upmu\mathrm{s}$\} and the fidelity is 98.3\%. In the experiment, the lattice beam is controlled by the accusto optical modulators (AOMs), which ensures the precise opening and closing of the optical lattice, thus achieving shortcut pulse control.

\begin{figure}[!htbp]
    \centering
    \includegraphics[width=0.6\linewidth]{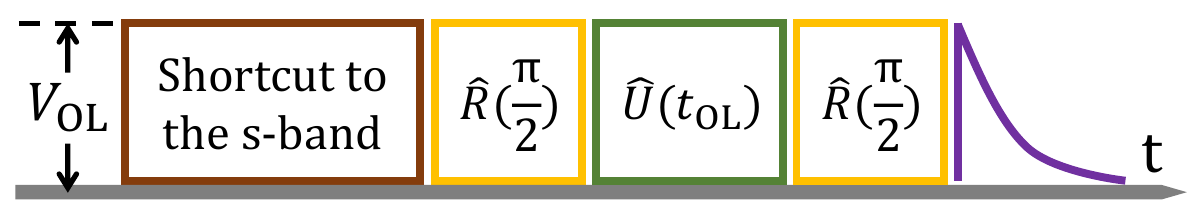}
    \caption{Time sequence for the RI. Condensates are loaded into the S-band initially by shortcut method, followed by the RI sequence: the first $\uppi/2$ pulse, evolution time $t_{\mathrm{OL}}$, and the second $\uppi/2$ pulse. The population of atoms in different bands are obtained after the \textit{band mapping} process.}
    \label{fig:fig2}
\end{figure}

The experimental process of the RI is shown in Fig.\ref{fig:fig2}. The first step is the shortcut to the S-band, which transfers BECs from the hybrid trap to the $\Gamma$ points of the S-band $\left|\psi_S\right>$ by a two-pulse shortcut sequence ($14.0/27.5/99.5/14.5\,\upmu\mathrm{s}$) with the fidelity of 99.3\%. Thus the initial states of the RI is $\psi_i=\left|\psi_{S}\right>$. Next, the first $\uppi/2$ pulse splits the atoms into two bands (S- and D-band) equally. Then, the triangular lattice will remain on for $t_{\mathrm{OL}}$, and the atoms in two bands interfere after the second $\uppi/2$ pulse. A \textit{band mapping} technique\cite{PhysRevLett.74.1542, PhysRevLett.87.160405} is utilized at the end of the sequence, with which we can get the distribution of atoms in different bands. During the band mapping process, the lattice potential is ramped down adiabatically in the form $\mathrm{e}^{-t/\tau_m}$, with the time constant $\tau_m=200\,\upmu\mathrm{s}$ for a total time of $1\,\mathrm{ms}$\cite{PhysRevA.104.033326}. Finally, we take absorption imaging with 31 $\mathrm{ms}$ time-of-flight (TOF) to obtain the distribution of atoms in different bands, such as the images in Fig.\ref{fig:fig3}(a).

\section{The Ramsey fringes in the triangular lattice}

\begin{figure}[!htbp]
	\centering
	\includegraphics[width=0.68\linewidth]{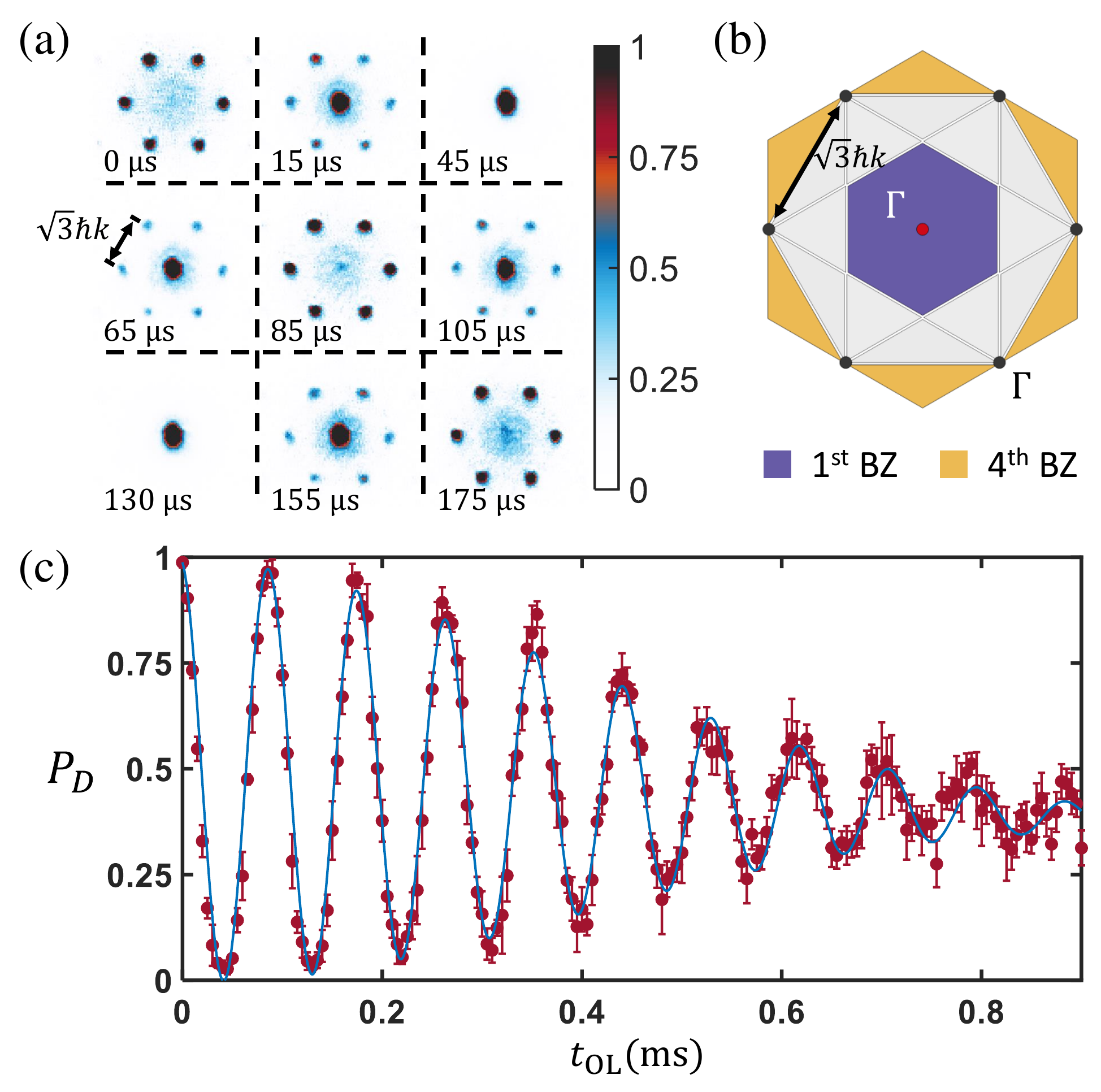}
	\caption{(a) Typical images at the initial holding time $t_{\mathrm{OL}}$. These pictures are taken after the \textit{mapping} technique with $31\,\mathrm{ms}$ time of fight. (b) Sketch of the Brillouin zone (BZ) of the triangular optical lattice. Blue area for the first BZ and yellow area for the 4th BZ. $\Gamma$ points for the first and the fourth BZ are marked as the red and the black point, respectively. (c) The oscillation of atoms populated in the D-band versus $t_{\mathrm{OL}}$. Each data point is the average of five measurements, and their standard deviation contributes to the errobar. The fitting curve is displayed with the blue solid line and a guide to eye. Codes for data analyzing are shown in Code file 1.}
	\label{fig:fig3}
\end{figure}\textbf{}

In the typical experimental images of Fig.\ref{fig:fig3}(a), the atoms are mainly distributed at seven points with an interval corresponding to a quais-momentum of $\sqrt{3}\hbar k$. Different positions on the images represent different bands. As shown in Fig.\ref{fig:fig3}(b), the blue and yellow areas represent the fisrt and fourth Brillouin zones (BZs), and the red and black points are the $\Gamma$ points of the S- and D-band, respectively. For example, in Fig.\ref{fig:fig3}(a), the atoms when $t_{\mathrm{OL}}=45\,\upmu\mathrm{s}$ are mainly distributed at the red point of Fig.\ref{fig:fig3}(b), which indicates the atoms are mainly populated at the $\Gamma$ point of the S-band. Similarly, the image for $t_{\mathrm{OL}}=0\,\upmu\mathrm{s}$ shows that the atoms are in the D-band.

By analyzing the absorption images with a bi-mode function\cite{Hu2018}, we can get the fractions of condensed atoms on the S- and D-band ($P_S,\,P_D$ respectively). We change the holding time $t_{\mathrm{OL}}$ and observe the oscillation of the $P_D$, as shown in Fig.\ref{fig:fig3}(c). For $t_{\mathrm{OL}}<0.175\,\mathrm{ms}$, the oscillation is very clear and contrast $C_R$ of the oscillation is close to 1, which is consistent with the theoretical expectation in Eq.(\ref{eq:ramsey}). This proves that we have successfully observed the Ramsey fringe and constructed the RI in the triangular lattice. The period of the oscillation is $T_c=88.8\,\upmu\mathrm{s}$, matching the band gap between the S- and D-band. The contrast $C_R$ is obtained by $(\max(P_D)-\min(P_D))/(\max(P_D)+\min(P_D))$ in one oscillation. For a longer holding time $t_{\mathrm{OL}}$, the amplitude of oscillation in Fig.\ref{fig:fig3}(c) will attenuate. The change of the contrast $C_R$ over holding time $t_{\mathrm{OL}}$ is shown in Fig.\ref{fig:fig4}(a). To demonstrate the performance of our RI, we define the coherence time $\tau$ as the time when the contrast $C_R$ drops to $1/\mathrm{e}$. The coherence time is $0.56\,\mathrm{ms}$ for the RI in the triangular optical lattice as shown in Fig.\ref{fig:fig4}(a).

\section{Analysis on decay mechanism of contrast}

\begin{figure}[!htbp]
	\centering
	\includegraphics[width=0.9\linewidth]{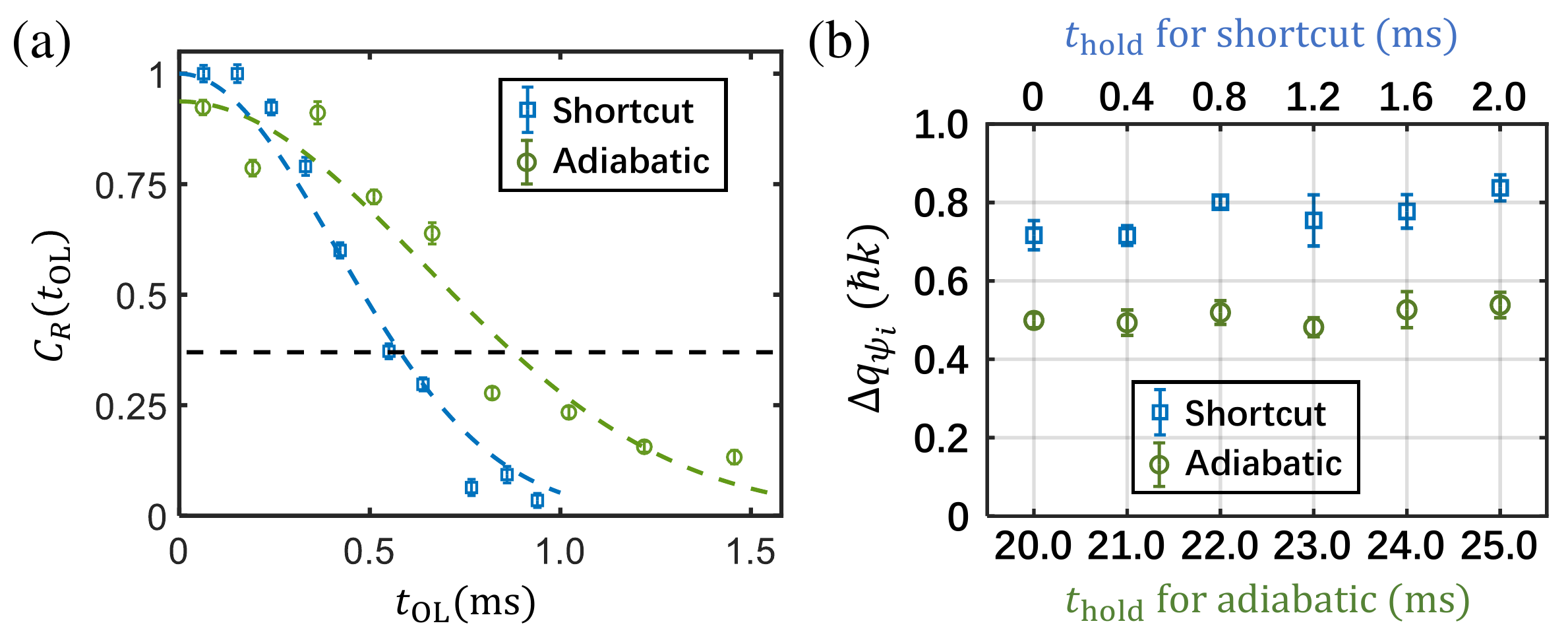}
	\caption{(a) The contrasts $C_R(t_\mathrm{OL})$ decay versus $t_\mathrm{OL}$ for RIs with shortcut (blue squares) and with adiabatic loading method (green circles). Dashed lines with the same color are the fitting curves. (b) Atoms momentum distribution of the interferometric initial state $\psi_i$ for different loading methods. Shortcut is measured in the time range of $0\sim2\,\mathrm{ms}$ (blue square) while the adiabatic way $20\sim25\,\mathrm{ms}$ (green circles).}
	\label{fig:fig4}
\end{figure}\textbf{}

To improve the performance of our RI, we need to analyze the decay mechanism, which limits the coherence time. We speculate that the decay of contrast may be mainly caused by shortcut pulse heating because of its non-adiabatical process. The consequence of heating is the broadening of quasi-momentum distribution, which will lead to dephasing. In order to confirm this conjecture, we carry out corresponding verifications in experiments and numerical simulation.

Firstly, we measure the broadening of the quasi-momentum distribution during the RI in the experiment. In the beginning, the BECs in the hybrid trap are transferred to the S-band by the shortcut pulse, and then the lattice is kept on for a holding time $t_\mathrm{hold}$. Next, the lattice is turned off abruptly and the images for atomic momentum distribution are obtained after the TOF process. By the bi-mode fitting\cite{Hu2018}, we can extract the width of the central momentum peak, which corresponds to the quasi-momentum distribution width of the atoms in the triangular optical lattice. The results are shown in Fig.\ref{fig:fig4}(b) with the blue squares. The initial distribution width is $\Delta q(t_\mathrm{hold}=0)=0.72\,\hbar k$, and it gradually gets wider as $t_\mathrm{hold}$ increases, reaches to a value close to $0.84\,\hbar k$ after $2\,\mathrm{ms}$. To better illustrate the broadening of quasi-momentum from the shortcut, we conduct a comparative experiment with an adiabatic method\cite{Jin_2019}, which causes very slight heating. The process is realized by an adiabatic rising-time of $80\,\mathrm{ms}$ followed by a pre-holding time of $t_\mathrm{hold}=20\,\mathrm{ms}$, which can also transfer atoms to the $\Gamma$ point of the S-band instead of the shortcut. For this adiabatic case, we measure the atomic quasi-momentum distributions from $t_\mathrm{hold}=20.0\,\mathrm{ms}$ to $25.0\,\mathrm{ms}$ and they all keep in a narrower range below $0.56\,\hbar k$, which is shown in Fig.\ref{fig:fig4}(b) with the green circles. The difference between the two results shows that the broadening of quasi-momentum (or the heating effect) caused by shortcut pulses does exist.

The wider distribution of quasi-momentum will cause dephasing, thus leading to the decay of contrast. In Fig.\ref{fig:fig1}(b), the band gap $\Delta\epsilon$ between S- and D-band is associated with the quasi-momentum $\vec{q}$. According to the Eq.(\ref{eq:ramsey}), for the non-zero quasi-momentum distribution, the Ramsey fringe is the superposition of different oscillations related to the $\Delta\epsilon(\vec{q})$. With the increase of time $t_\mathrm{OL}$, the phases of different oscillations will be disorderly, resulting in contrast attenuation, which is called dephasing. With the experimental results of the quasi-momentum distribution, we can further analyze the decay mechanism by numerical simulation. With a multi-band simulation method\cite{PhysRevA.104.L060601} considering the quasi-momentum distribution, we obtain the simulation Ramsey fringe of the RI. Similar to the previous experiments, we extract the coherence time of $0.63\,\mathrm{ms}$, which is slightly longer than the experimental results $0.56\,\mathrm{ms}$. This is because the RI also includes other decay mechanisms not considered in the simulation. For example, the heating of the $\uppi/2$ pulse and the decay caused by the non-uniform beams.

In order to verify our theoretical analysis, we carry out a comparative experiment. We replace part of the pulse sequence in Fig.\ref{fig:fig2} with the adiabatic way to attenuate the broadening effect. Then a new experimental scheme for RI in the triangular lattice is proposed: the preparation stage for the interferometric initial state $\psi_i=\left|\psi_S\right>$ is realized by the adiabatic method. The interference fringe contrasts $C_R$ of the RI with adiabatic loading are plotted in Fig.\ref{fig:fig4}(a). The coherence time for adiabatic RI is $\tau=0.88\,\mathrm{ms}$, which is consistent with our simulation result of $0.96\,\mathrm{ms}$. This coherence time is nearly 2 times longer than that of the shortcut, which indicates that the dephasing caused by the quasi-momentum distribution is the main reason for the decay of the contrast.

\section{An echo-Ramsey interferometer}

In the last section, we introduce the adiabatic RI, which extends the coherence time. However, this method needs a very long manipulation time, which is not suitable for some scenarios requiring rapid control\cite{PhysRevA.104.L060601}. Therefore, we need to search for fast methods to improve the coherence time. From our analysis, we know that the dephasing contributes the most to the decay of the contrast, which can usually be eliminated by the echo technology\cite{Hu2018}. Hence, we further develope a band echo $\uppi$ pulse technique with the shortcut method. The $\uppi$ pulse can exchange the atoms population and reverse the phase disorder, hence inhibiting some decoherence mechanisms, for example, the dephasing in our RI\cite{Hu2018}. The design principle of $\uppi$ pulse sequence is similar to that of the $\uppi/2$ pulse. The final states $\psi'_{f1}=\hat{R}(\uppi)\left|\psi_S\right>$ and $\psi'_{f2}=\hat{R}(\uppi)\left|\psi_D\right>$ should satisfy the conditions $\psi'_{f1}=\left|\psi'_{m1}\right>\equiv\left|\psi_D\right>$ and  $\psi'_{f2}=\left|\psi'_{m2}\right>\equiv-\left|\psi_S\right>$. And the fidelity is $\eta'=\left|\left<\psi'_{f1}|\psi'_{m1}\right>+\left<\psi'_{f2}|\psi'_{m2}\right>\right|/2$. We get the time sequence for the $\uppi$ pulse \{$(t_1^{\rm{on}},t_1^{\rm{off}},t_2^{\rm{on}},t_2^{\rm{off}})=(34.0,6.5,9.0,78.0)\,\upmu\mathrm{s}$\} with the fidelity of 93.0\%.

\begin{figure}[!htbp]
	\centering
	\includegraphics[width=0.9\linewidth]{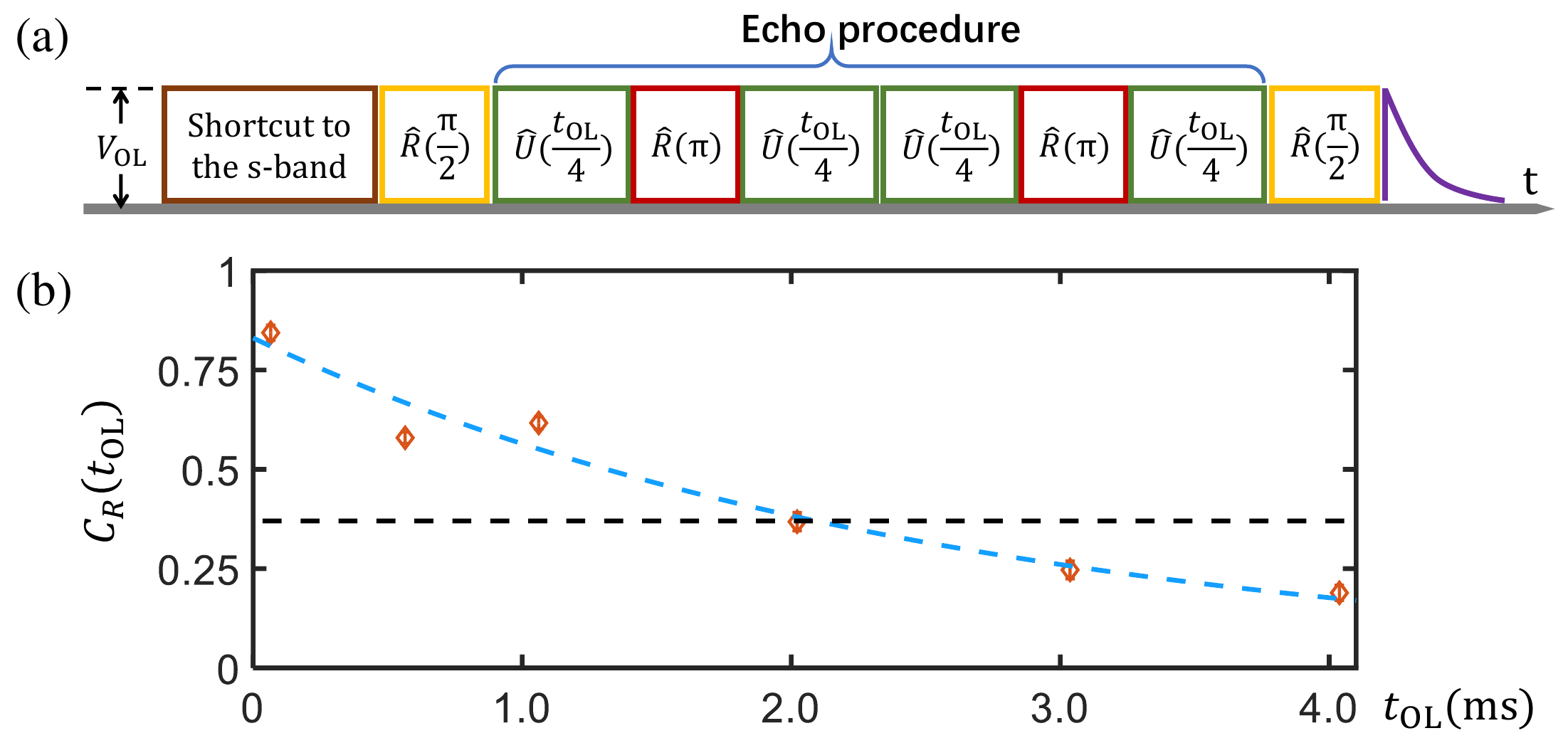}
	\caption{(a) Time sequence for an echo-RI. Two $\uppi$ pulses are inserted between the $\uppi/2$ pulses. (b) The contrast $C_R(t_\mathrm{OL})$ decay versus $t_\mathrm{OL}$ for a $2\times\uppi$ echo-RI experimental scheme. The brown rhombuses are experimental data points while the blue dashed line is the fitting curves.}
	\label{fig:fig5} 
\end{figure}

Then we construct the echo-Ramsey interferometer with two $\uppi$ pulses, which is depicted in Fig.\ref{fig:fig5}(a), leading to a final state of
\begin{equation}
	\psi'_{f}=\hat{R}(\uppi/2)[\hat{U}(t_{OL/4})\hat{R}(\uppi)\hat{U}(t_{OL/4})]^2\hat{R}(\uppi/2)\psi_i,
\end{equation}
where the holding times before and after a $\uppi$ pulse are equal to each other for satisfying the condition of echo process\cite{Hu2018}.

\textbf{}

The contrast $C_R(t_\mathrm{OL})$ decay of a $2\times\uppi$ echo-RI is illustrated in Fig.\ref{fig:fig5}(b) and we get a coherence time of $\tau=2.11\,\mathrm{ms}$ by fitting curve, which indicates the significant improvement of the coherence time. It should be noted that the first data in Fig.\ref{fig:fig5}(b) is not 1: only about 80\% of the atoms have been transferred to the D-band. This is mainly caused by the low fidelity of the shortcut $\uppi$ pulse (93\%), which is far lower than that of $\uppi$/2 pulse. This reason is that the rotation angle of the $\uppi$ pulse on the Bloch sphere is $\uppi$, larger than the $\uppi$/2 pulse, which means the longer manipulation trace. However, if we improve the shortcut method, the fidelity can also be very high. For example, if we design a pulse sequence with variable amplitude \{$(t_1^{\rm{on}},t_1^{\rm{off}},...,t_5^{\rm{on}},t_5^{\rm{off}})=(18.1,11.8,5.7,39.7,0.3,20.5,11.8,0.1,85.9,10.7)\,\upmu\mathrm{s}$\} and the lattice depth for each pulse \{$(3.6,5.2,4.3,5.1,3.9)\,\mathrm{E_r}$\}, we can achieve 99.2\% fidelity, which could be realized in the future experiment after the technology update.

\section{Discussion and conclusion}
We successfully design the $\uppi/2$ and $\uppi$ pulse and realize the RI in this complex two-dimensionally coupled optical lattice for the first time. Compared with the 1D lattice in our previous experiments\cite{Hu2018}, the coherence time in the triangular lattice is shorter. This is because the broadening of the quasi-momentum distribution is more obvious in the triangular lattice. If we assume that the quasi-momentum broadening of the 2D lattice is the same as that of the 1D lattice, which is $0.20\,\hbar k$, then the 2D coherence time obtained by numerical simulation will reach $6.3\,\mathrm{ms}$, significantly higher than that in 1D potential ($1.4\,\mathrm{ms}$). Codes for simulations are shown in Code file 2.

The performance of this RI still has the potential for further improvement after overcoming some technological limitations in current experiments. The first is the alignment of the triangular lattice. We observe a position drift of atoms when processing the alignments with a deeper lattice depth, which implies that the lattice beams do not precisely intersect at the same point (also should be the position of BECs). We attribute the imperfect alignments to the stability of the optical path, which reduces the coherence time of RI. In our system of 2D lattice, no fiber cable is applied before entering the vacuum chamber. A fiber cable can circumvent this problem by converting optical direction changes into power jitters. The second limitation is depth stability. Because the rise and fall time of the $\uppi/2$ and $\uppi$ shortcut pulses should be less than 100 ns, it is difficult to construct a proper power feedback loop. The power jitter of light intensity will cause the attenuation of Ramsey fringes. In the future, a fast feedback loop can be applied to stabilize the power of lattice beams and improve the fidelity of shortcut pulses by introducing more variable parameters, such as the amplitudes of each pulse. Third, non-uniform potential distribution in the radial of the Gaussian beam increases the dephasing effect\cite{Hu2018}, which can be weakened by using apodizing filters. Next, the phases between the lattice lights can be locked, making the lattice more stable and reducing thermal effects. Finally, composite lattices can be used to construct flat bands to reduce the dephasing effect and improve the coherence time. In these complex lattices, we can construct a similar RI by designing appropriate $\uppi/2$ and $\uppi$ pulses because the shortcut methods are still applicable.

In the future, this RI is expected to be applied to unveil quantum many-body dynamics in optical lattices, quantum information, and precision measurement\cite{Hu2018}. For example, the direct application of the interferometer is to evaluate the coherence time of the atom-orbital qubit in optical lattices. In the previous work\cite{PhysRevA.104.L060601}, the interferometer was used to measure the coherence time of a single qubit in a 1D optical lattice. To construct multiple qubits with atom-orbital, 2D or 3D optical lattices are necessary. The RI for the two-dimensional lattice in this paper will contribute to the construction of multiple qubits based on atom-orbital in the future. We can analyze the decoherence mechanisms according to Ramsey fringes\cite{Hu2018}, to improve the fidelity and operability time of the qubits in 2D lattices. In addition, we can use this RI to measure the bands of optical lattices. The corresponding $\uppi$/2 pulses are designed for quantum states at different quasi-momentum to construct the Ramsey interferometer. The frequency of the Ramsey fringes corresponds to the band gap. Then we can use it to measure the band distortion caused by interaction\cite{PhysRevLett.118.175301} or topological characteristics\cite{Tarnowski2019} in 2D optical lattices.

In summary, we have demonstrated a Ramsey interferometer in the two-dimensionally coupled triangular optical lattice. By means of shortcut, a $\uppi/2$ pulse between the S- and D-band is obtained, based on which we observe the time-dependent Ramsey fringes. The fringes have got an oscillation period of $88.8\,\upmu\mathrm{s}$ and a coherence time of $0.56\,\mathrm{ms}$. After careful analysis in comparative experiments and numerical simulations, we find that the dephasing casued by the quasi-momentum distribution of atoms contributes the most to the decay of the fringes. For this decay mechanism, we develope the echo $\uppi$ pulse technique and significant increase the coherence time. These deliberate quantum control technologies in such dimensionally coupled system could contribute to the areas of quantum information and precise measurements based on external quantum states of atoms.

\section{Backmatter}

\begin{backmatter}
\bmsection{Funding}
This work is supported by the National Basic Research Program of China (Grants No. 2021YFA0718300 and No. 2021YFA1400901), the National Natural Science Foundation of China (Grants No. 12104020, No. 61727819, No. 12004360, and No. 11334001), the Science and Technology Major Project of Shanxi (No. 202101030201022), and the Space Application System of China Manned Space Program.

\bmsection{Disclosures}
The authors declare no conflicts of interest.

\bmsection{Data Availability} Data underlying the results presented in this paper are available in Code file 1, Ref.\cite{code_ana}, and Code file 2, Ref.\cite{code_sim}.

\end{backmatter}

\bibliography{Ref.bib}
\end{document}